# Hunting for Massive Black Holes in Dwarf Galaxies

*Amy E. Reines*
*eXtreme Gravity Institute, Montana State University*

**Despite traditional thinking, an appreciable population of massive black holes may be lurking in dwarf galaxies. Prior to the last decade, nearly all massive black holes were found in the nuclei of giant galaxies and the existence of massive black holes in dwarf galaxies was highly controversial. The field has now been transformed with a growing community of researchers working on a variety of observational studies and theoretical models of dwarf galaxies hosting massive black holes. Work in this area is not only important for a holistic understanding of dwarf galaxy evolution and feedback, but it may just tell us how the first "seeds" of massive black holes formed in the early Universe. In this Perspective, I discuss the current state of the field as well as future prospects. I also present new insights on the demographics of nearby dwarf galaxies, which can be used to help constrain the black hole occupation/active fraction as a function of mass and dwarf galaxy type.**

Astronomers have long theorized about the origin of supermassive black holes (BHs) with typical masses of $M_{\rm BH} \sim 10^6 - 10^9\ M_\odot$ [ref. 1]. These BHs inhabit the nuclei of virtually all massive galaxies including our Milky Way, and they are the engines that power active galactic nuclei (AGNs) from local Seyferts to high-redshift quasars[2]. We also now know that at least some dwarf galaxies host BHs with $M_{\rm BH} \lesssim 10^6\ M_\odot$ [ref. 3]. Such BHs are called by many names: intermediate-mass BHs (i.e., between stellar-mass and supermassive BHs), low-mass BHs (i.e., compared to typical supermassive BHs), or massive BHs (i.e., more massive than stellar-mass BHs). I will use "massive BH" or simply "BH" throughout this Perspective. Dwarf galaxies are typically defined in this context to have stellar masses similar to the Large Magellanic Cloud or lower[3] with $M_{\rm stellar} \lesssim 10^{9.5}\ M_\odot$ (or up to $10^{10}\ M_\odot$ in some studies[4]) and all massive BH detections in dwarf galaxies thus far have been in "bright dwarfs"[5] with $M_{\rm stellar} \gtrsim 10^7\ M_\odot$ (although there is a recent claim of a BH in the dwarf spheroidal Leo I[6]).

BHs in dwarf galaxies have not grown much compared to their more massive counterparts since dwarfs have relatively quiet merger histories that tend to deprive the BHs of favorable conditions to accrete significant amounts of gas[7]. Supernova feedback is also thought to hinder BH growth in dwarf galaxies[8,9]. Therefore, observations of BHs in dwarf galaxies place the most tangible constraints on the masses of BH "seeds" and provide clues to their formation mechanism[10].

The discovery of quasars with billion solar-mass BHs at redshifts when the Universe was less than a Gyr old tells us that the first massive BHs formed at very early times. However, unlike today's stellar-mass BHs that are recognized as the remnants of massive stars at the end-point of stellar evolution, we do not have a clear picture for the formation of the first massive BHs[10-12]. The seeds of massive BHs may have been prevalent in the early Universe, formed from the remnants of the first generation of (Population III) stars with light BH masses ($\lesssim 10^2\ M_\odot$) [ref.

13]. While such BHs almost certainly existed, it would be difficult for them to accrete enough material to become significantly more massive, and thus alternative scenarios have been proposed. For example, the conditions in dense star clusters could facilitate the formation of BH seeds with masses in the range of $\sim 10^2 - 10^4\ M_\odot$ [refs. 14-17]. Or perhaps the seeds were even heavier and rarer, formed in an exotic scenario where giant dense gas clouds formed supermassive stars that ultimately collapsed to form BH seeds with $\sim 10^5 M_\odot$ [refs. 12,18,19].

Dwarf galaxies are expected to host the smallest massive BHs and help distinguish between various models for seed formation. Indeed, observed relationships between BH mass and host galaxy stellar mass[20] show that the least massive BHs live in dwarf galaxies with the lightest BHs having masses $< 10^5\ M_\odot$ [refs. 3,21,22]. Moreover, models of BH growth over billions of years predict that the observational signatures of the dominant seeding mechanism at early times should be manifested in the present-day population of dwarf galaxies. For example, since we expect light seeds to be common and heavy seeds to be rare, the fraction of today's dwarf galaxies containing a massive BH (i.e., the occupation fraction) should help elucidate the seed formation mechanism at early times[7]. The low-mass end of scaling relations such as the $M_{\mathrm{BH}} - \sigma_\star$ relation may also help differentiate seeding models[23], although the subsequent BH accretion mode may obfuscate seeding signatures[24]. In any case, finding the least-massive BHs in today's dwarf galaxies gives us concrete (upper) limits on the masses of BH seeds. While the minimum mass of BHs in dwarf galaxies and the occupation fraction provide two independent constraints on BH seeding models, subsequent BH accretion and dynamical perturbations/ejections may complicate interpreting observational results.

Searching for and studying massive BHs in dwarf galaxies is also important for observationally constraining the impact of BH feedback at low masses. Feedback is an important ingredient in galaxy formation models, and necessary to match the predicted dark matter halo mass function to the observed galaxy luminosity function. AGN feedback is thought to regulate and suppress star formation in massive galaxies, while stellar feedback (e.g., supernovae and radiation from young host stars) is thought to dominate in low mass galaxies[25]. However, motivated by recent observational results, more studies are focusing on the role of BH feedback in dwarf galaxies[26,27]. We now know that massive BHs do exist in at least some dwarf galaxies, which has implications for the interplay between BH growth and star formation in these systems.

**Focus of this Perspective**

This Perspective is part of a *Nature Astronomy* special collection on dwarf galaxies. Given the emphasis on dwarf galaxies (as opposed to BHs) and the scope of a Nature Perspective, the focus here will be somewhat different than that of a recent comprehensive review on intermediate-mass BHs[28]. I will tend to highlight work explicitly studying BHs in "normal" intact dwarf galaxies with $M_{\mathrm{stellar}} \lesssim 10^{9.5}\ M_\odot$. However, massive BHs have also been found in compact stellar systems in this stellar mass range, including ultracompact dwarfs (UCDs) and compact ellipticals (cEs). While a handful of BHs with $M_{\mathrm{BH}} \sim 10^{6-7}\ M_\odot$ have been detected in UCDs[29-32] and cEs[33] via stellar dynamics, the relatively large BH-to-stellar mass ratios add to the evidence that these galaxies are stripped remnants of larger progenitor galaxies[34,35]. It is also possible that BHs exist in the stripped cores of accreted dwarf satellite galaxies (e.g., HLX-1 [ref. 36]) and may be discovered as "wandering" BHs in the outskirts of massive galaxies[37,38]. However, such BHs have yet to be discovered in the Milky Way halo[39].

**A brief history**

Not long ago, the very existence of massive BHs in dwarf galaxies was debated. Prior to the last decade, we only knew of a few oddballs including M32 [ref. 33] and the Seyfert 1 nuclei in NGC 4395 [ref. 40] (Figure 1) and Pox 52 [ref. 41]. A dozen or so additional examples of AGNs in dwarf galaxies were subsequently found in the Sloan Digital Sky Survey (SDSS), most of which were included in large samples of broad-line AGNs with $M_{\rm BH} < 2 \times 10^6 \, M_\odot$ [refs. 42-44], and a complementary sample of narrow-line AGNs in galaxies with relatively low velocity dispersions[45]. For the most part, however, these samples of AGNs do not probe the dwarf galaxy regime[45,46]. A few examples of low-level accretion powered massive BHs in early-type dwarf galaxies were also found using sensitive, high-resolution X-ray observations with the *Chandra X-ray Observatory*[47], and it was about a decade ago that the radio-detected highly sub-Eddington massive BH was discovered in the dwarf starburst galaxy Henize 2-10 [ref. 48-51] (Figure 1).

A larger population of BHs in dwarf galaxies remained elusive until 2013 when we found more than a hundred dwarf galaxies ($M_{\rm stellar} \sim 10^{8.5} - 10^{9.5} \, M_\odot$) in the SDSS with optical spectroscopic signatures of AGNs[3]. This systematic search demonstrated that massive BHs in dwarf galaxies were more common than previously realized. BH masses for the small number of broad-line AGNs detected were in the range of $M_{\rm BH} \sim 10^5 - 10^6 \, M_\odot$ and the active fraction based on narrow emission line ratios for the entire sample was found to be ~0.5%. However, the true BH occupation fraction is likely much higher since the objects found in this work represent just the tip of the iceberg. Other BHs that are inactive or have low accretion rates, as well as BHs residing in dwarf galaxies with significant ongoing star formation were undoubtedly missed.

There has since been a growing number of other searches for AGNs in dwarf galaxies using a variety of multi-wavelength tracers, as well as many follow-up studies of existing samples. In addition to considerable observational work on this topic, there have also been many recent theoretical investigations of massive BHs in dwarf galaxies using simulations[26,27,52-54] and semi-analytic models[24]. Below I focus on recent progress on the observational front.

**Finding accreting massive black holes in dwarf galaxies**

Reliably identifying massive BHs in dwarf galaxies presents some unique challenges relative to massive galaxies. First, while using stellar or gas dynamics to weigh a BH is generally considered the most secure method, applying this technique to even the nearest dwarf galaxies within the Local Group is at the edge of our current observational capabilities[22] since the BH radius of influence is too small to resolve in more distant dwarf galaxies. For example, the radius of influence of $10^5$ $M_\odot$ BH in a dwarf galaxy with a central stellar velocity dispersion of 30 km s$^{-1}$ is only ~0.5 pc, where $r_{\rm infl} = GM_{\rm BH}/\sigma^2$.

Even active BHs in dwarf galaxies are difficult to find and distinguish from star-formation-related emission. AGNs in dwarf galaxies are powered by smaller BHs than typical AGNs in more massive galaxies, and will therefore be less luminous at a given Eddington fraction. Therefore, we cannot use conventional luminosity limits and must lower our threshold for what might be an AGN in a dwarf galaxy. The consequence is that we must be vigilant and always consider the possibility of interlopers related to star formation. Additionally, nuclear star clusters are common in galaxies with stellar masses between $M_{\rm stellar} \sim 10^8 - 10^{10}$ $M_\odot$ and known to co-exist with massive BHs in many galaxies with $M_{\rm stellar} \sim 10^9 - 10^{10}$ $M_\odot$ [ref. 55]. The presence of these clusters may further complicate detecting massive BHs in dwarf galaxies through dynamics or accretion signatures.

*Optical Selection:*

The most common approaches to identify AGNs using optical spectroscopy include searching for narrow emission line photoionization signatures of active BHs and/or searching for broad H$\alpha$ emission that could signify dense gas orbiting within the gravitational potential of a massive BH. While narrow-line diagnostics can be used to identify the presence of massive BHs, broad-line AGNs have the added benefit of yielding BH mass estimates using the width and luminosity of the broad line.

In the low-mass regime, some studies begin with a sample of dwarf galaxies (typically using a stellar mass cut) and then search for AGN signatures[3,4,56], while other studies first search for broad-line AGNs and then select objects with relatively low BH masses[42,44,57]. In the latter case, not all of the host galaxies are necessarily dwarfs. Also, since broad-line AGNs are harder to detect than narrow-line AGNs for low-mass BHs, the number of bona-fide dwarf galaxies is usually small in these samples. Moreover, follow-up studies demonstrate that broad H$\alpha$ alone should not be taken as conclusive evidence for an AGN in a dwarf galaxy since supernovae can also make (short-lived) broad H$\alpha$[58]. Narrow-line ratios indicating the presence of an AGN are also key. Observations with the *Chandra X-ray Observatory* have confirmed the presence of active massive BHs in some dwarf galaxies, and the Eddington ratios for these optically-selected AGNs are in the range of $\sim 0.1\% - 50\%$ [ref. 59].

The dwarf host galaxies with spectroscopically-selected AGNs have stellar masses comparable to the Magellanic Clouds[3], but for the most part do not look like the Magellanic Clouds. An *HST* study[60] of 41 dwarf host galaxies found that the majority (85%) have regular morphologies, and most of these are disk-dominated with small pseudobulges. A handful of irregular galaxies were also found (15% of the observed sample) including Magellanic-types and disturbed galaxies showing signs of interactions/mergers. An analysis of the stellar populations[61] indicate the hosts are old dwarf galaxies with stellar mass-weighted ages in the range of $10^9 - 10^{10}$ yrs. Compared to the general population of dwarf galaxies in the same mass range, the optically-selected AGN-hosting dwarfs have a significantly lower fraction of young stellar populations contributing to the stellar mass[61].

It should be emphasized that samples of active massive BHs in dwarf galaxies selected via optical spectroscopy as described above are severely biased. These searches will not detect BHs that are highly obscured, weakly accreting at very low Eddington fractions, or those living in galaxies with lots of ongoing star formation. We know that BHs with high accretion rates are rare[62] and that dwarf galaxies are predominantly star forming[63], so we are almost certainly missing the bulk of the population using single-fiber optical spectroscopic searches alone (i.e., from the SDSS). Additionally, AGNs in low-metallicity galaxies, which dwarfs tend to be, can overlap with low-metallicity starburst galaxies in traditional emission line diagnostic diagrams and may be missed[64]. X-ray binary populations are also enhanced in low-metallicity galaxies, which can affect emission lines (e.g., stronger He II)[65] and possibly mimic AGNs.

Other approaches in the optical regime, including photometric variability[66-69] and spatially-resolved spectroscopy[70-73], can help alleviate some of the bias against finding AGNs in predominantly star-forming dwarf galaxies. The detection of optical [FeX] coronal line emission has also recently been used to find a sizeable sample of massive BHs in dwarf galaxies largely missed by traditional optical searches[74]. Interestingly, the [FeX] luminosities are consistent with accretion onto massive BHs from either AGNs or tidal disruption events (TDEs). Indeed, there is already strong evidence for TDEs in some dwarf galaxies[75,76].

A variety of other AGN tracers spanning the electromagnetic spectrum have been employed to search for additional examples of massive BHs in dwarf galaxies. While many of these AGN candidates in dwarf galaxies appear to be robust, interlopers related to star formation are almost always a concern in the low-mass / low-luminosity regime (even more so than at optical wavelengths). Background AGNs and mismatched sources also need to be considered. Below I focus on what we have learned from a number of studies using radio, X-ray, and mid-infrared observations.

*Radio:*

Radio continuum observations have been used to identify candidate AGNs in dwarf galaxies[48,49,77-81] via synchrotron emission from AGN jets (or the base of a compact jet). The detection of circumnuclear water maser emission at GHz frequencies also offers the opportunity to detect and accurately measure masses of low-mass BHs[82]. In general, radio continuum observations (e.g., with the Very Large Array; VLA) require much less time than X-ray observations (e.g., with *Chandra*) so a sensible strategy to construct a sample of massive BHs is to start with a radio survey and follow up candidates with X-rays. Using radio observations alone, we must be cognizant of potential interlopers that can masquerade as AGNs including supernova remnants and younger supernovae that produce synchrotron emission, as well as dense HII regions that emit thermal bremsstrahlung[81]. I discuss two recent systematic searches for radio AGNs in dwarf galaxies below.

The first study[80] presents a sample of 35 radio-selected AGN candidates culled from the VLA-COSMOS 3 GHz Large Project. The host galaxies were selected to have stellar masses $\lesssim 10^{9.5}$ $M_\odot$ and the global radio luminosities are above that expected from star formation. This sample also extends to higher redshifts than most searches ($z \sim 3.4$). However, reliably disentangling the contributions from star-formation-related emission is particularly thorny at these large distances and the redshifts themselves primarily come from photometric techniques, adding additional uncertainty to galaxy stellar masses and radio luminosities.

The other study[81] used the VLA to target a sizeable sample (~100) of nearby ($z < 0.055$) dwarf galaxies and search for luminous, compact radio emission associated with AGN jets. After considering various possible origins for the compact (~0.25" or 85 pc at the median distance of the sample) radio sources detected with their VLA observations, a final sample of 13 dwarf galaxies with compelling AGN candidates was found. One object was previously identified as an optical AGN[3], and another has recently been confirmed as an AGN via the coronal [Fe X] line using Gemini integral field spectroscopy[73]. The latter work also finds evidence for an outflow and shocked gas, suggesting BH feedback is important in radio-selected AGNs in dwarf galaxies. The sample of 13 dwarf galaxies consists of both nuclear and non-nuclear radio-selected BHs[81]. While background interlopers still need to be definitively ruled out for the non-nuclear sources, there is tantalizing evidence that at least some of the off-nuclear radio sources are indeed "wandering" massive BHs associated with the target dwarf galaxies. First, the dwarf galaxies with radio-selected AGNs have systematically higher ratios of [OI]/H$\alpha$ relative to the other dwarfs that have less luminous radio sources consistent with star formation. Second, the more extended/disturbed dwarf galaxies tend to have more offset radio sources, consistent with a scenario in which the BHs have been displaced due to galaxy mergers[52,53].

While both radio-selected samples discussed above consist of bluer and lower-mass galaxies than optical searches, they are biased towards the most luminous radio sources. Both studies impose a radio pre-selection criterion. In one study[80], the galaxies must be classified as AGNs in the VLA-COSMOS 3 GHz Large Project. In the other[81], the target dwarf galaxies were pre-selected to have detections in the Faint Images of the Radio Sky at Twenty centimeters (FIRST) survey. Therefore, these works could be missing a significant population of massive BHs in dwarf galaxies that do not produce radio continuum emission at the levels that are detectable in these surveys.

*X-ray:*

X-ray observations of dwarf galaxies offer another approach to detect massive BHs via their accretion flows[47,50,83-90]. X-rays can also probe massive BHs with lower Eddington ratios than those typically picked up by optical searches. However, we must be wary of individual stellar-mass X-ray binaries (XRBs) with comparable luminosities, as well as their aggregate emission in more distant galaxies. Background AGNs are also a concern when the X-ray source is significantly offset from the galaxy center, as is the case for some of the detected objects[84,87,88].

Samples of X-ray selected AGNs in dwarf galaxies are typically found by cross-matching a mass-selected sample of dwarf galaxies with archival X-ray observations or existing X-ray catalogs derived from pointed X-ray observations (e.g., with *Chandra* or *XMM*). The expected galaxy-wide contribution from XRBs is generally estimated using galaxy star formation rates and stellar masses[91], and there is also a dependence on metallicity[89,92]. A significant excess X-ray luminosity is often taken as evidence for an AGN, although extra caution should be applied in the case of dwarf galaxies since the scaling relations are not well constrained at low masses and stochastic effects can lead to a small number of luminous XRBs dominating the total X-ray luminosity. On the other hand, relatively dormant massive BHs need not produce luminous X-ray emission and may remain undetected.

*Mid-IR:*

There have also been numerous attempts to select AGNs in dwarf galaxies using mid-IR color diagnostics obtained from the *Wide-field Infrared Survey Explorer (WISE)*[56,93-95]. However, careful analyses[89,96,97] have demonstrated that these samples are unreliable and likely full of contaminants. In luminous AGNs, the accreting BH heats surrounding dust that emits a red power-law spectrum in the mid-IR. However, it has been shown that dwarf starburst galaxies can heat dust in such a way that mimics the mid-infrared colors of AGNs, and that a single mid-infrared color cut is insufficient to robustly select AGNs in dwarf galaxies[96]. Subsequent work[89] that incorporated *Chandra* and *HST* observations found that even a more stringent two-color selection criterion fails to reliably identify AGNs in optically-star-forming dwarf galaxies. Moreover, the majority of optically-selected AGNs in dwarf galaxies are dominated by host galaxy light in mid-infrared color space with only a small fraction showing up as mid-infrared selected AGNs[96]. Finally, the relatively poor resolution of *WISE* compared to optical surveys can result in wrongly associating infrared AGNs with dwarf galaxies when there are in fact multiple optical sources (including more luminous galaxies) overlapping with one *WISE* source[97]. In sum, it is clear that attempting to select mid-infrared AGNs in dwarf galaxies at the resolution of *WISE* is problematic.

**Implications for black hole seeding and feedback in dwarf galaxies**

The fraction of dwarf galaxies hosting massive BHs is important for our general understanding of BH demographics, dwarf galaxy evolution, and a key diagnostic for discriminating between various BH seed formation scenarios at early times. The studies described above find AGN fractions in dwarf galaxies that are typically $\lesssim 1\%$. However, this only places a lower limit on the BH occupation fraction since not all BHs will produce the requisite accretion signatures to be detected.

The best constraints on the BH occupation fraction at low mass are in regular galaxies with $M_{\text{stellar}} = 10^9 - 10^{10}\ M_\odot$ and come from dynamics and X-ray surveys. Dynamical methods, which rely on detecting the BH's gravitational influence on nearby stars or gas, suggest an occupation fraction of ~80% for early-type galaxies in this stellar mass range[22]. The BH occupation fraction in galaxies with $M_{\text{stellar}} = 10^9 - 10^{10}\ M_\odot$ drops to > 50% when including late-type spirals[28], since most of these have upper limits on BH masses[98]. In all, there are only ten galaxies in this mass range with published dynamical BH masses or limits[28], so these values should be treated with caution. *Chandra* observations of nearby galaxies in the volume-limited AMUSE surveys suggest an occupation fraction > 20% [ref. 85]. At first glance, these results point toward a heavy seeding model when compared to semi-analytic models[24]. However, we stress that these results are relevant for early-type dwarf galaxies, which are not representative of the overall dwarf galaxy population. The AMUSE X-ray surveys target early-type galaxies where star formation is practically absent to help alleviate contamination from high-mass XRBs[47].

In Box 1, I investigate the demographics of nearby dwarf galaxies so we can more accurately interpret constraints on the BH occupation fraction. Taking a recent version of the Catalog and Atlas of the Local Volume Galaxies[99] (updated 2021 Aug 17) as a representative sample of galaxies, and assuming the reported *K*-band luminosities (in $L_\odot$) are good proxies for stellar mass (in $M_\odot$, at least to first order[100]), I find that early-type galaxies with morphological *T* types ≤ 1 constitute only ~13% of galaxies with $M_{\text{stellar}} = 10^9 - 10^{10}\ M_\odot$. Similar results are found by integrating stellar mass functions derived from the Galaxy and Mass Assembly Survey of the local Universe[101] (0.025 < *z* < 0.06) - elliptical galaxies constitute 11% of all galaxies with $M_{\text{stellar}} = 10^9 - 10^{10}\ M_\odot$. While there is complexity in morphologically classifying galaxies and making comparisons between various samples, it is clear that we should be cautious interpreting results regarding BH seeding scenarios based exclusively on the fraction of early-type dwarf galaxies hosting BHs.

We currently do not know if the BH occupation fractions are different for early-type and late-type dwarf galaxies, but we do know that the demographics of the dwarf galaxies themselves differ substantially (Box 1). On the one hand, the occupation fraction in early-type dwarfs may be skewed high given the association between galaxy spheroids and BHs[2], in which case a heavy seeding model may be preferred since this would lead to an overall low BH occupation fraction in dwarfs. On the other hand, we know very little about the demographics of BHs in late-type dwarfs, which make up the vast majority of the dwarf galaxy population. There are hints from X-ray observations that find AGN candidates in ~20% and ~10% of late-type Sd-Sdm and Sm-Im galaxies, respectively, with higher fractions in the barred variety of these galaxies[102]. The galaxy stellar masses of this sample, while not given in that work, are likely to be ≲ $10^{10} M_\odot$ (Figure 2). Using the Local Volume Galaxy sample[99] again, I find that late-type galaxies with morphological *T* types ≥ 7 constitute ~72% of galaxies with $M_{\text{stellar}} = 10^9 - 10^{10}\ M_\odot$. We currently have no meaningful constraints on the BH occupation fraction in irregular and blue compact dwarf (BCD) galaxies, although there are at least a handful of detections[48,78,103]. These types of dwarfs dominate the galaxy population at stellar masses $M_{\text{stellar}} = 10^7 - 10^{8.5}\ M_\odot$ (Table 1), but finding massive BHs in these galaxies is especially challenging given they are primarily star forming with shallow potential wells and the BHs may be non-nuclear[52,79,81]. Nevertheless, if massive BHs are ubiquitous in late-type/irregular dwarf galaxies (i.e., the most common types of dwarfs), a light seeding scenario may be favored since this would push the overall BH occupation fraction at low masses towards unity.

Scaling relations between BHs and their host galaxy properties at low mass may also provide clues to the origin of massive BHs. For example, the slope and scatter of the low-mass end of the $M_{\rm BH} - \sigma_\star$ relation may differ depending on whether BH seeds were light or heavy[23], although recent work suggests there is also a dependence on the accretion mode that can hinder our ability to discriminate between seeding models[24]. In any case, dwarf galaxies appear to broadly follow the same $M_{\rm BH} - \sigma_\star$ [ref. 28,104] and $M_{\rm BH} - M_{\rm bulge}$ [ref. 105] relations as higher mass systems, albeit with larger scatter. For the most part, dwarf galaxies also share the same BH mass to total stellar mass relation ($M_{\rm BH} - M_{\rm stellar}$) as spiral galaxies, which has a significantly lower normalization than the relation for elliptical/early-type galaxies[20,28]. Moreover, given that the lowest mass BHs have $M_{\rm BH} < 10^5\ M_\odot$, we can rule out seeding scenarios that exclusively produce BHs with larger masses. It is worth noting that the majority of BH mass estimates in dwarf galaxies are indirect and come from broad-line AGNs. The masses carry large uncertainties (~0.5 dex) due to assumptions regarding gas virialization, the geometry of the broad-line region, and extrapolating quasar scaling relations to low-mass BHs[106]. Using spectra from a flux limited survey such as the SDSS will also result in an observational bias toward finding the most massive and luminous BHs in dwarf galaxies with detectable broad H$\alpha$ emission (e.g., see Section 3.6 in ref. 3). Deeper spectroscopic surveys with higher spectral resolution are needed to discover broad-line AGNs with lower BH masses[21].

Regarding the impact of massive BHs on dwarf galaxies, there is a growing body of evidence suggesting BH feedback is at play in the low-mass regime. AGN-driven outflows have been discovered in dwarf galaxies with optically-selected[107,108] and radio-selected AGNs[51,73] using spatially resolved kinematics. Additionally, a study of HI masses finds that some of the dwarf galaxies with optical signatures of active BHs are significantly gas-depleted compared to the general population of galaxies in the same stellar mass range[109]. Environmental processes are ruled out since all of the galaxies under consideration are isolated. While most studies focus on negative AGN feedback, which suppresses star formation, a new study presents the first example of positive BH feedback in a dwarf galaxy[51]. Using *HST* spectroscopy of the central regions in Henize 2-10, we provide evidence for BH-triggered star formation including the detection of a quasi-linear ionized filament with a velocity structure that is well-described by a simple precessing bipolar outflow model. All of these studies support the idea that galaxy formation models should include BH feedback in dwarf galaxies.

Indeed, recent observational progress has motivated more theorists to explore BH activity in dwarf galaxies. Analytical work suggests that negative AGN feedback may be more effective than supernovae at clearing gas from early dwarf galaxies[110], and could potentially explain a number of dwarf galaxy anomalies that challenge our understanding of galaxy formation[111]. Numerical simulations yield mixed results. Some studies indicate strong supernova feedback can stunt BH growth at low masses[8,9]. While these results reaffirm dwarf galaxies are a good testbed for BH seed formation, they also suggest AGN feedback at low masses should be largely absent. However, other numerical studies find that AGN feedback can produce significant outflows and suppress/quench star formation in dwarf galaxies[26,27,112].

**Future observations**

Our understanding of the demographics and impact of massive BHs in dwarf galaxies is far from complete, but as described above we have made a lot of progress over the last decade. Upcoming multiwavelength observational capabilities are sure to advance the field going forward with a number of new avenues to pursue both AGNs and quiescent BHs in dwarf galaxies. These studies will be important for determining the active/occupation fraction of BHs in dwarf galaxies, as well as the role of BH feedback in these systems.

In terms of sheer numbers, finding and studying active BHs will continue to dominate in the dwarf galaxy regime. In the next few years, the Vera C. Rubin Observatory will begin to carry out the 10-yr Legacy Survey of Space and Time (LSST) that will explore the transient optical sky and accelerate variability searches for AGNs in dwarf galaxies[66-69]. TDE detections with the Rubin Observatory are also very promising for BH demographics measurements both in terms of finding and weighing low-mass BHs[113] and in terms of constraining the occupation fraction[114]. The Square Kilometer Array (SKA) and a next generation VLA (ngVLA) could play prominent roles in detecting optically elusive massive BHs in dwarf galaxies with their unprecedented sensitivity and resolution at radio wavelengths[115]. At high energies, *eROSITA* will complete an all-sky X-ray survey that may produce more than a thousand AGN candidates in nearby dwarf galaxies[90]. Additionally, future observations with *Lynx* could provide deep X-ray observations at high angular resolution, which are crucial for identifying AGNs in dwarf galaxies both near and far. In the infrared, the *James Webb Space Telescope* (*JWST*) will give us sensitive, high resolution images that will help detect and disentangle AGN emission from the dwarf host galaxies, which is currently not possible with *WISE*. *JWST* spectroscopy will also enable the study of infrared coronal emission lines associated with AGNs in dwarf galaxies[116-119]. At high redshifts, *JWST* color-color selection will offer the opportunity to detect and separate AGNs from star-formation-dominated galaxies[120] and help discriminate between initial seeding mechanisms[121].

Beyond accretion signatures, there are also reasons to be optimistic about increasing the number of dynamical searches for massive BHs in dwarf galaxies in the coming years. With current telescopes, the sphere of influence of a $10^5$ $M_\odot$ BH in a dwarf galaxy can only be resolved out to distances of $\sim 1$ Mpc. Upcoming 30-meter class telescopes will be able to push to $\sim 10$ Mpc and greatly expand the accessible volume in which dynamical searches can be performed. There are hundreds of dwarf galaxies within this volume[99] and the next generation of giant ground-based telescopes will be sensitive to quiescent BHs that are currently out of reach. Such studies could greatly improve constraints on the BH occupation fraction at low masses as a function of different types of dwarf galaxies (Box 1, Table 1).

Looking further ahead toward the era of the *Laser Interferometry Space Antenna (LISA)*, we may be in for a giant leap in our understanding of the demographics of massive BHs in dwarf galaxies. *LISA* will detect gravitational waves from merging BHs with masses in the range $\sim 10^4 - 10^7$ $M_\odot$ at redshifts up to $\sim 20$, and cosmological simulations predict[52] that BH mergers in dwarf galaxies occur throughout cosmic time from $z \sim 12$ to the present.

Gravitational wave detections also have the potential to place strong constraints on the formation mechanism of BH seeds[24,122]. However, the merger rates will also depend on BH dynamics which may complicate interpreting the results[54].

A slew of new observatories and large astronomical surveys will no doubt be a boon to our efforts to find and study massive BHs in dwarf galaxies in the next decade and beyond. Combined with advances in modelling and simulations, these observations will provide new insights into BH seed formation and growth, as well as the impact of BH feedback in dwarf galaxies. Plus, if the past decade is any indicator, we are likely in store for some unexpected surprises.

**Box 1 – The demographics of dwarf galaxies as a means to constrain the black hole occupation fraction**

Dwarf galaxies come in many different flavors and we must understand their demographics if we are to constrain the occupation/active fraction of massive BHs at low mass. To this end, I investigate the demographics of dwarf galaxies using the most recent version of the Catalog and Atlas of the Local Volume Galaxies[99]. There are ~1000 galaxies with $K$-band luminosities less than the LMC (log $L_K/L_\odot$ < 9.42) in the catalog, the majority of which have distances within ~11 Mpc. Figure 2 shows morphology/dwarf galaxy type as a function of luminosity, and Figure 3 shows the number distributions of various types of dwarf galaxies as a function of luminosity. Table 1 shows the fraction of different types of dwarf galaxies in various luminosity bins and is meant to assist researchers in constraining the occupation/active fraction of massive BHs in the low mass regime.

There are two important caveats to note. First, the local volume used here may not be fully representative of the Universe since there are no massive galaxy clusters included. Second, while $K$-band luminosities (in $L_\odot$) are decent proxies for stellar mass (in $M_\odot$) to first order, $M/L_K$ ratios can vary by a factor of ~3 across early and late-type galaxies at low masses[100].

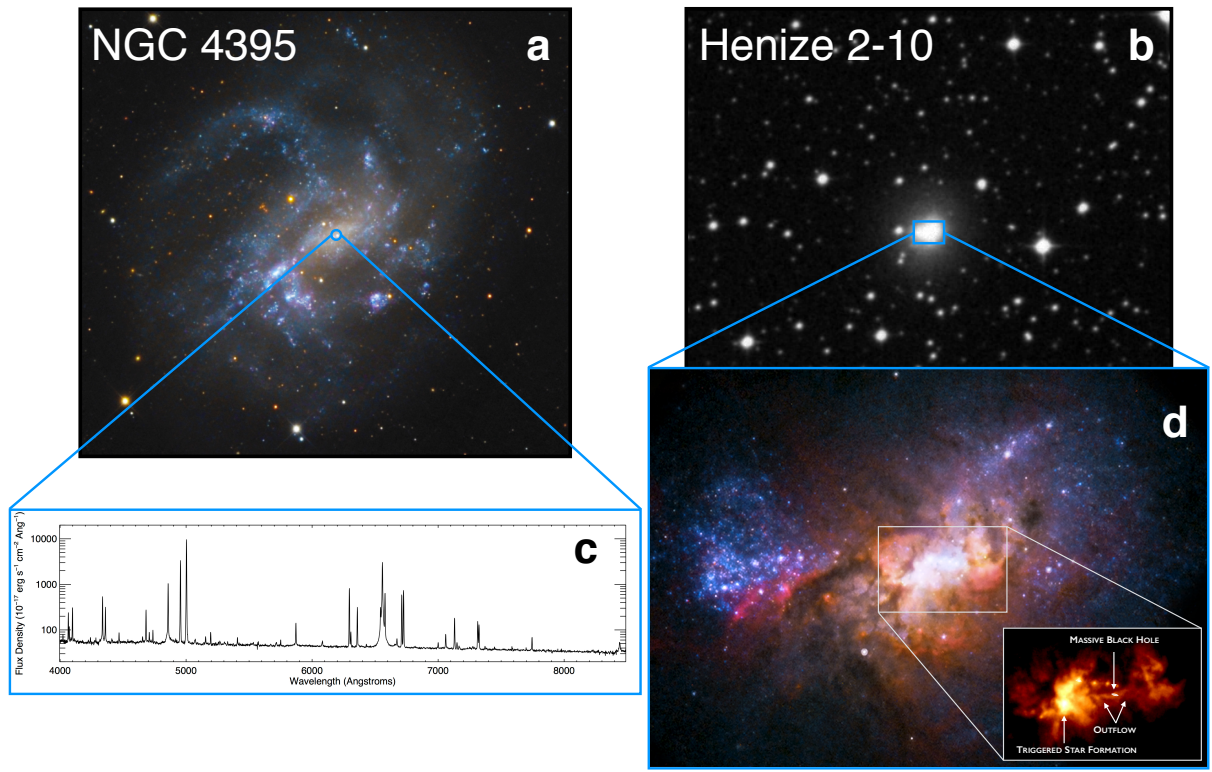

**Figure 1: The dwarf galaxy NGC 4395 has an optically-selected AGN, while Henize 2-10 has a radio-selected highly sub-Eddington massive black hole. a,** Optical image of NGC 4395 with a field of view of 15 arcminutes (∼17.5 kpc at a distance of 4 Mpc). Image courtesy of Bob Franke/Focal Point Observatory. **b,** Digital Sky Survey image of Henize 2-10 with a field of view of 6.6 arcminutes (i.e., the same physical scale as the NGC 4395 image; ∼17.5 kpc at a distance of 9 Mpc). **c,** Optical spectrum of the nucleus of NGC 4395 from the Sloan Digital Sky Survey. The spectrum exhibits AGN-like narrow emission line ratios, as well as broad Balmer lines from gas orbiting within the potential of the massive black hole. **d,** Hubble Space Telescope optical image of Henize 2-10 with a field of view of 25 arcseconds (∼1.1 kpc). Image credit: NASA/STScI. The inset shows narrowband Hα+continuum imaging and radio contours from Very Long Baseline Interferometry in white, with the radio detection of the massive black hole indicated[49]. The black hole is also detected in X-rays[50]. Recent work shows that the black hole is driving a bipolar outflow that is triggering star formation in the central region of the galaxy[51]. Image from Schutte & Reines 2021.

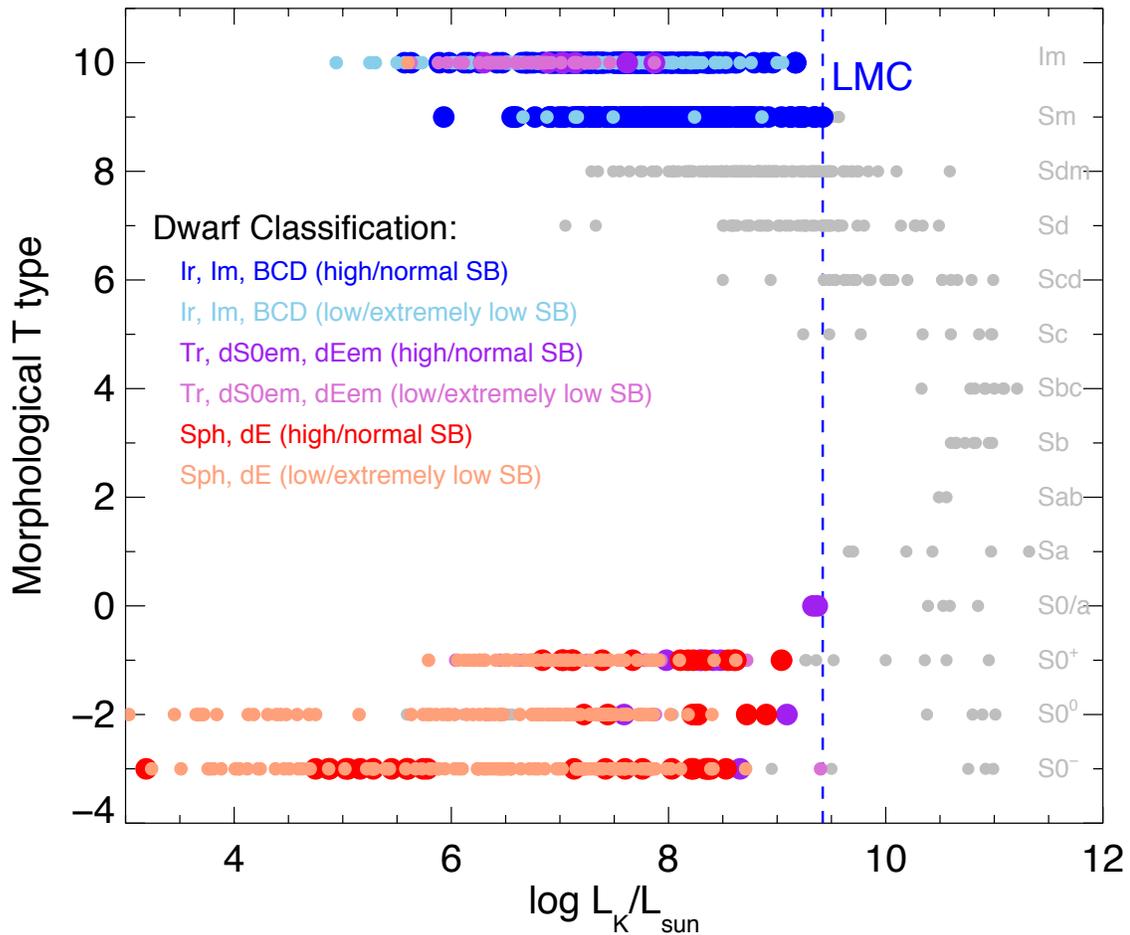

**Figure 2: Morphology versus log K-band luminosity for 1245 galaxies**. The data are taken from the Catalog and Atlas of the Local Volume Galaxies[99]. To first order, the *K*-band luminosity (in $L_\odot$) is a good proxy for stellar mass (in $M_\odot$) [ref. 100]. Morphological T types are based on the classification by de Vaucouleurs et al. (1991)[123]. Colored points indicate dwarf galaxies fainter than the LMC that are further classified[99] according to their surface brightness (high, normal, low or extremely low) and color (or the presence of emission). High/normal surface brightness galaxies are indicated by large points and low/extremely low surface brightness galaxies are indicated by small points. The vertical dished line marks the luminosity of the LMC. Blue dwarf galaxies tend to be high surface brightness blue compact dwarfs (BCDs) or lower surface brightness irregulars (Ir,Im). Red dwarf galaxies tend to be high surface brightness dwarf ellipticals (dEs) or low surface brightness dwarf spheroidals (Sph). "Transient" type dwarf galaxies (Tr) combine properties of both spheroidals and irregulars and have mixed colors. Gray points indicate galaxies that have not been given a more detailed morphological classification. The vast majority of these that are fainter than the LMC are late-type spirals (Sdm and Sd).

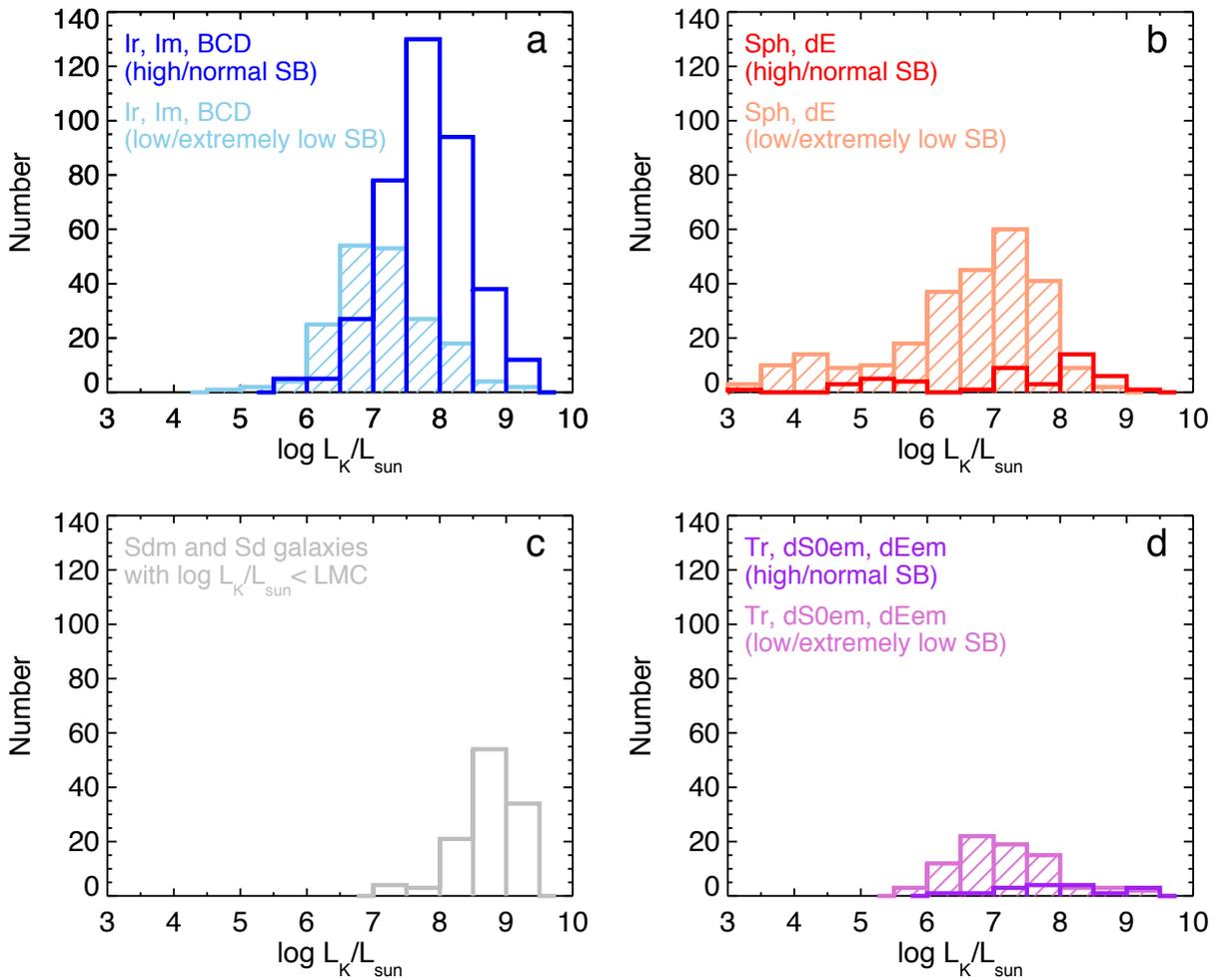

**Figure 3: Panels show the number distributions of various types of dwarf galaxies fainter than the LMC (log $L_K/L_\odot$ < 9.42) as a function of K-band luminosity.** The data are taken from the Catalog and Atlas of the Local Volume Galaxies[99]. **a**, The distribution for Irregulars, Magellanic Irregulars and Blue Compact Dwarf Galaxies. The dark blue histogram includes galaxies with high or normal surface brightness and the light blue histogram includes those with low or extremely low surface brightness. **b**, The distribution for spheroidals and dwarf ellipticals. The red histogram includes galaxies with high or normal surface brightness and the light peach histogram includes those with low or extremely low surface brightness. **c**, The distribution for late-type spiral galaxies (Sdm and Sd) with K-band luminosities less than the LMC. **d**, The distribution for transient type dwarf galaxies, and dwarf lenticulars and ellipticals with emission lines. The dark purple histogram includes galaxies with high or normal surface brightness and the light purple histogram includes those with low or extremely low surface brightness.

|  | 4.0 - 4.5 | 4.5 - 5.0 | 5.0 - 5.5 | 5.5 - 6.0 | 6.0 - 6.5 | 6.5 - 7.0 | 7.0 - 7.5 | 7.5 - 8.0 | 8.0 - 8.5 | 8.5 - 9.0 | 9.0 - 9.5 |
|---|---|---|---|---|---|---|---|---|---|---|---|
| Sdm, Sd | 0 | 0 | 0 | 0 | 0 | 0 | 0.02 | 0.01 | 0.13 | 0.50 | 0.63 |
| Ir, Im, BCD | 0 | 0 | 0 | 0.15 | 0.06 | 0.18 | 0.35 | 0.58 | 0.58 | 0.35 | 0.22 |
| Ir, Im, BCD | 0 | 0.08 | 0.12 | 0.12 | 0.31 | 0.36 | 0.23 | 0.12 | 0.11 | 0.04 | 0.04 |
| Tr, dS0em, dEem | 0 | 0 | 0 | 0 | 0.01 | 0.01 | 0.01 | 0.02 | 0.02 | 0.01 | 0.06 |
| Tr, dS0em, dEem | 0 | 0 | 0 | 0.09 | 0.15 | 0.15 | 0.08 | 0.07 | 0.02 | 0.03 | 0.04 |
| Sph, dE | 0 | 0.23 | 0.29 | 0.12 | 0 | 0.01 | 0.04 | 0.01 | 0.09 | 0.06 | 0.02 |
| Sph, dE | 1 | 0.69 | 0.59 | 0.53 | 0.46 | 0.30 | 0.27 | 0.18 | 0.06 | 0.02 | 0 |

**Table 1**: **The fraction of dwarf galaxies in various luminosity bins are derived from the Catalog and Atlas of the Local Volume Galaxies**. The data are taken from the Catalog and Atlas of the Local Volume Galaxies[99]. Luminosity bins given in bold in the top row are in units of log $L_K/L_\odot$. These values can be used to help constrain the BH occupation fraction in dwarf galaxies by accounting for the galaxy type and luminosity/mass probed by various studies.

**Data Availability:**
The data that support the findings of this study are available in The Astronomical Journal with the identifier 10.1088/0004-6256/145/4/101 [ref. 99]. Correspondence and requests for other materials should be addressed A.E.R.

**Acknowledgments:**
A.E.R acknowledges support for this work provided by Montana State University and NASA through EPSCoR grant number 80NSSC20M0231.

**Author contributions:**
A.E.R carried out the analysis and wrote the paper.

**Competing interests:**
The author declares no competing interests.